\documentclass[a4paper]{article}
\usepackage{Odyssey2018}
\usepackage{etex} 

\usepackage{calc}
\usepackage{color}
\usepackage{xcolor} 
\usepackage{preambles/colorbrewer}
\usepackage{placeins} 

\usepackage{graphicx}
\usepackage[utf8]{inputenc}
\makeatletter
	\@ifpackageloaded{babel}{}{\usepackage[ngerman,german,american]{babel}}
\makeatother

\usepackage[shortcuts]{extdash}

\usepackage{longtable}
\usepackage{booktabs,array,tabularx}
\usepackage{multirow,arydshln}
\usepackage{amssymb,amsmath,bm,mathtools,MnSymbol}
\usepackage{listings}

\usepackage[final]{pdfpages}

\usepackage[caption=false,font=footnotesize,subrefformat=parens,labelformat=parens]{subfig}

\newcolumntype{L}[1]{>{\hsize=#1\hsize\raggedright\arraybackslash}X}%
\newcolumntype{R}[1]{>{\hsize=#1\hsize\raggedleft\arraybackslash}X}%
\newcolumntype{C}[1]{>{\hsize=#1\hsize\centering\arraybackslash}X}%
\newcolumntype{P}[1]{>{\centering\arraybackslash}p{#1}}

\usepackage{todonotes} 
\usepackage{verbatim} 
\usepackage{xifthen} 

\makeatletter\newcommand*{\missingreference}[1]{\colorbox{red}{?#1?}}
\newcommand*{\missingcitation}[1]{\colorbox{red}{?#1?}}
\makeatletter
\def\@setref#1#2#3{%
	\ifx#1\relax
	\protect\G@refundefinedtrue
	\nfss@text{\reset@font\missingreference{#3}}
	\@latex@warning{Reference `#3' on page \thepage \space
		undefined}%
	\else
	\expandafter#2#1\null
	\fi}
\def\@citex[#1]#2{\leavevmode
	\let\@citea\@empty
	\@cite{\@for\@citeb:=#2\do
		{\@citea\def\@citea{,\penalty\@m\ }%
			\edef\@citeb{\expandafter\@firstofone\@citeb\@empty}%
			\if@filesw\immediate\write\@auxout{\string\citation{\@citeb}}\fi
			\@ifundefined{b@\@citeb}{\hbox{\reset@font\missingcitation{#2}}
				\G@refundefinedtrue
				\@latex@warning
				{Citation `\@citeb' on page \thepage \space undefined}}%
			{\@cite@ofmt{\csname b@\@citeb\endcsname}}}}{#1}}
\makeatother

\usepackage{ulem}
\usepackage{tikz}
\usepackage{pgfplots}
\pgfplotsset{compat=newest}

\usetikzlibrary{patterns,shapes,arrows,decorations}
\usetikzlibrary{calc,shadings,patterns}
\usetikzlibrary{automata,chains}
\usetikzlibrary{shadows,calc,backgrounds}
\usepgflibrary{shapes.geometric,shapes.symbols}
\usepgfplotslibrary{fillbetween,polar,smithchart}
\usepgfplotslibrary{colorbrewer}

\pgfmathdeclarefunction{gauss}{2}{%
	\pgfmathparse{1/(#2*sqrt(2*pi))*exp(-((x-#1)^2)/(2*#2^2))}%
}
\pgfmathdeclarefunction{erf}{1}{%
    \begingroup
    \pgfmathparse{#1 > 0 ? 1 : -1}%
    \edef\sign{\pgfmathresult}%
    \pgfmathparse{abs(#1)}%
    \edef\x{\pgfmathresult}%
    \pgfmathparse{1/(1+0.3275911*\x)}%
    \edef\t{\pgfmathresult}%
    \pgfmathparse{%
        1 - (((((1.061405429*\t -1.453152027)*\t) + 1.421413741)*\t 
        -0.284496736)*\t + 0.254829592)*\t*exp(-(\x*\x))}%
    \edef\y{\pgfmathresult}%
    \pgfmathparse{(\sign)*\y}%
    \pgfmath@smuggleone\pgfmathresult%
    \endgroup
}
\tikzset{
    declare function={inverf(\x)=\x/abs(\x) * sqrt( sqrt( (4.3307 + ln(1-\x^2)/2 )^2 - ln(1-\x^2)/0.147 ) - (4.3307 + ln(1-\x^2)/2);},
    declare function={probit(\p)=1.4142 * inverf(2*\p-1);}
}
\pgfmathdeclarefunction{logit}{1}{%
    \pgfmathparse{ln(#1/(1-#1))}%
}
\pgfmathdeclarefunction{logitpdf}{3}{
\pgfmathparse{exp(-((#1-#2)/#3))/(s*(1+exp(-((#1-#2)/#3)))^2)}%
}
\pgfmathdeclarefunction{logitcdf}{3}{
\pgfmathparse{1/(1+exp(-((#1-#2)/#3)))}%
}

\tikzset{ 
	every plot/.style={prefix=plots/pgf-}, 
	shape dasec/.style={ 
		color=hda, 
		draw,
		fill=white,
		line width=0.01cm,
		minimum height=1.25em,
		inner xsep=0.0cm,
		inner ysep=0.0cm
	},
    shape database/.style={
        cylinder,
        cylinder uses custom fill,
        cylinder body fill=white!50,
        cylinder end fill=white!50,
        shape border rotate=90,
        aspect=0.25,
        draw=hda,
        text=hda
    }
}

\def\addlegendimage{\csname pgfplots@addlegendimage\endcsname}

\usepackage{cmap}
\usetikzlibrary{shapes.misc}
\usetikzlibrary{mindmap}

\tikzset{
    module step/.style={anchor=west,align=left,fill=white,inner sep=0.25ex,draw=blue!25,text width=\stepwidth},
    module data/.style={align=center,fill=white,inner sep=0.25ex,draw=blue!25,text width=\datawidth},
    channel/.style={anchor=west,align=left,fill=white,inner sep=0.25ex,draw=blue!25,text width=\commwidth},
    pics/module/.style n args={3}{code={
            \begin{scope}[on background layer]%
                \path [rounded corners,top color=blue!10, bottom color=blue!25, draw=blue!25] %
                ([xshift=-2ex,yshift=1.5ex]#2.north west) %
                node [rounded corners=0pt,inner sep=0.25ex,draw=orange,fill=white,right=1ex] %
                {#1} %
                rectangle ([xshift=2ex,yshift=-.75ex]#3.south east);%
            \end{scope}%
    }},
    pics/channel left/.style n args={7}{code={%
            \begin{scope}[on background layer]
                \node [single arrow, drop shadow, shape border uses incircle, shape border rotate=-180,
                thick, draw=orange!75, inner color=orange!20, outer color=orange!35,
                minimum height=#4, minimum width=#5, anchor=center, single arrow head extend=#6, single arrow tip angle=#7]
                at ([xshift=#2,yshift=#3]#1) {};
            \end{scope}
    }},
    pics/channel right/.style n args={7}{code={%
    \begin{scope}[on background layer]
        \node [single arrow, drop shadow, shape border uses incircle, shape border rotate=0,
        thick, draw=orange!75, inner color=orange!20, outer color=orange!35,
        minimum height=#4, minimum width=#5, anchor=center, single arrow head extend=#6, single arrow tip angle=#7]
        at ([xshift=#2,yshift=#3]#1) {};
    \end{scope}
    }},
    pics/channel down/.style n args={7}{code={%
            \begin{scope}[on background layer]
                \node [single arrow, drop shadow, shape border uses incircle, shape border rotate=-90,
                thick, draw=orange!75, inner color=orange!20, outer color=orange!35,
                minimum height=#4, minimum width=#5, anchor=center, single arrow head extend=#6, single arrow tip angle=#7]
                at ([yshift=#2,xshift=#3]#1) {};
            \end{scope}
    }}
}

\newcommand{\setvariable}[2]{
	\let#1\relax
	\newcommand{#1}{#2}
}

\let\oldensuremath\ensuremath
\renewcommand{\ensuremath}[1]{
    \relax\ifmmode
    \oldensuremath{#1}
    \else
    \oldensuremath{#1}\@\xspace
    \fi
}

\usepackage{xspace} 
\usepackage[use-xspace,alsoload=prefixed,alsoload=abbr,alsoload=binary]{siunitx}
\newcommand{\bits}{\ensuremath{\,\mathrm{bits}}}

\newcommand{\KiB}{\ensuremath{\,\mathrm{KiB}}}

\newcommand{\MiB}{\ensuremath{\,\mathrm{MiB}}}

\newcommand*{\eg}{e.g.\@\xspace}

\newcommand*{\ie}{i.e.\@\xspace}

\newcommand*{\cf}{cf.\@\xspace}
\newcommand*{\etal}{et al.\@\xspace}
\makeatletter
\newcommand*{\etc}{%
	\@ifnextchar{.}%
	{etc}%
	{etc.\@\xspace}%
}
\makeatother

\allowdisplaybreaks

\DeclareMathOperator{\lcm}{lcm}

\DeclareMathOperator{\Z}{\mathbb{Z}}

\NewDocumentCommand{\ceil}{s O{} m}{%
	\IfBooleanTF{#1} 
	{\left\lceil#3\right\rceil} 
	{#2\lceil#3#2\rceil} 
}
\NewDocumentCommand{\floor}{s O{} m}{%
	\IfBooleanTF{#1} 
	{\left\lfloor#3\right\rfloor} 
	{#2\lfloor#3#2\rfloor} 
}

\newcommand{\Cllr}   {\ensuremath{C_{\textrm{llr}}}}
\newcommand{\Cllrmin}{\ensuremath{C^{\textrm{min}}_{\textrm{llr}}}}

\DeclareMathOperator{\enc}{enc}
\DeclareMathOperator{\dec}{dec}

\definecolor{hda}{rgb}{0.64,0,0.15}
\definecolor{mycolor1}{rgb}{0.00000,0.44700,0.74100} 
\xdefinecolor{mightyslate}{rgb}{0.33,0.38,0.44}
\xdefinecolor{pacifica}{rgb}{0.305,0.804,0.769}
\xdefinecolor{applicchic}{rgb}{0.78,0.957,0.392}
\xdefinecolor{cherrypink}{HTML}{9D98FF}
\xdefinecolor{grandmaspillow}{rgb}{0.769,0.302,0.345}
\definecolor{darkgreen}{rgb}{0.2,0.7,0.3}
\definecolor{darkorange}{rgb}{1,0.55,0}

\PassOptionsToPackage{hyphens}{url}\usepackage[bookmarks=false,hidelinks]{hyperref} 

\ninept

\title{Homomorphic Encryption for Speaker Recognition: \\ Protection of Biometric Templates and Vendor Model Parameters}

\makeatletter
\def\name#1{\gdef\@name{#1\\}}
\makeatother
\name{{\em Andreas Nautsch, Sergey Isadskiy, Jascha Kolberg, Marta Gomez-Barrero, Christoph Busch}}

\address{da/sec -- Biometrics and Internet Security Research Group, 
    Hochschule Darmstadt, Germany  \\
    {\small \tt \{andreas.nautsch,sergey.isadskiy,jascha.kolberg\}@h-da.de} \\
    {\small \tt \{marta.gomez-barrero,christoph.busch\}@h-da.de
}}

\setvariable\stepwidth{6em}
\setvariable\commwidth{5em}
\setvariable\datawidth{5em}
\newcommand{\matharchitecturenode}[1]{#1}
\newcommand{\stepno}[1]{\ensuremath{\textbf{\color{hda}#1}}}

\begin{document}
\maketitle

\begin{abstract}
Data privacy is crucial when dealing with biometric data.
Accounting for the latest European data privacy regulation and payment service directive, biometric template protection is essential for any commercial application.
Ensuring \emph{unlinkability} across biometric service operators, \emph{irreversibility} of leaked encrypted templates, and \emph{renewability} of \eg, voice models following the i\-/vector paradigm, biometric voice-based systems are prepared for the latest EU data privacy legislation.
Employing Paillier cryptosystems, Euclidean and cosine comparators are known to ensure data privacy demands, without loss of discrimination nor calibration performance.
Bridging gaps from template protection to speaker recognition, two architectures are proposed for the two\-/covariance comparator, serving as a generative model in this study.
The first architecture preserves privacy of biometric data capture subjects.
In the second architecture, model parameters of the comparator are encrypted as well, such that biometric service providers can supply the same comparison modules employing different key pairs to multiple biometric service operators.
An experimental proof\-/of\-/concept and complexity analysis is carried out on the data from the 2013\,--\,2014 NIST i\-/vector machine learning challenge.
\end{abstract}

\section{Introduction}
\label{sec:introduction}

The latest EU data privacy regulation \cite{EU-Directive-DataPrivacy-160427} declares biometric information as \emph{personal data}, \ie highly sensitive and entitled to the right of privacy preservation. Similarly, the current payment service directive \cite{EU-Directive-PSD2-151125} also requires biometric template protection to be employed in biometric systems utilized for banking services. To that end, the ISO/IEC IS~24745 \cite{ISO-IEC-24745-TemplateProtection-101129} on biometric information protection provides guidance on how to preserve the subject's privacy by defining the following three main properties to be fulfilled by protected biometric templates:
\begin{itemize}
    \item \emph{unlinkability}: stored biometric templates shall not be linkable across applications or databases,
    \item \emph{irreversibility}: biometric samples cannot be reconstructed from protected biometric templates,
    \item \emph{renewability}: multiple biometric references can be independently transformed, when they are created from one or more samples of a biometric data capture subject.
\end{itemize}
In addition to these properties, other performance metrics, such as recognition accuracy, should be preserved.

Even if some works argue that there is no need for template protection depending on the feature extraction \cite{Vaquero-VoiceBTP-Interspeech-2015}, sensitive information can be derived from unprotected templates, as it has already been proved for other biometric characteristics~\cite{Cappelli07PAMIReconstruction,Galbally2013}. In particular, linkability of state\-/of\-/the\-/art speaker recognition features is demonstrated in \cite{Glembek-MigratingIvectors-Interspeech-2015} with the motivation of interchanging features among different voice biometric services.
The interchange of biometric data across services is recently addressed ethically in \cite{Moyakine-SIIPPanelEthics-EABrpc-2017}, especially when targeting forensic and investigatory scenarios. Accounting for latest data privacy legislations, we motivate template protection, especially in commercial but also in other dual\-/use case application scenarios.


Current approaches to biometric template protection can be broadly classified into three categories \cite{Rathgeb-BTP-Survey-EURASIP-2011}, namely: $i)$ cancelable biometrics \cite{patel15CancelableBioSurvey}, where irreversible transformations are applied at sample or template level; $ii)$ cryptobiometric systems \cite{campisi13secPrivacyBio}, where a key is either bound or extracted from the biometric data; and $iii)$ biometrics in the encrypted domain \cite{fontaine13surveyHE}, where techniques based on homomorphic encryption (HE) and garbled circuits are used to protect the data. 
Whereas cancelable biometrics and cryptobiometric systems usually report some accuracy degradation \cite{Rathgeb-BTP-Survey-EURASIP-2011}, the use of HE schemes prevents such loss, since the operations carried out in the encrypted domain are equivalent to those performed with plaintext data. For this reason, we apply in this work HE schemes similar to the ones proposed in \cite{gomez2016implementation,GomezBarrero-BiometricTemplateProtection-InformationScience-2016,GomezBarrero-MBTPwithHE-PR-2017,GomezBarrero-VariableLengthBTPHE-Access-2017} to speaker recognition relying on generative comparators, such as probabilistic linear discriminant analysis (PLDA). We thereby ensure data privacy for data capture subjects for comparison models utilizing the two\-/covariance (2Cov) approach \cite{Cumani-PairwiseDiscriminativeVerification-TASLP-2013,Cumani-GenerativePairwiseModels-Odyssey-2014} (\ie full subspace PLDA) as a prototype generative comparison algorithm.

In contrast to conventional discriminative comparators, generative models can emit features with associated likelihoods based on pre\-/trained models.
Thus, comparison scores of generative models represent probabilistic similarity.
In this context, supplying model parameters to various service operators can arise privacy concerns regarding the data protection of biometric service vendors, \ie the pre\-/trained models.
Therefore, we further propose a mutual encryption scheme granting subject and vendor data privacy by employing well\-/established Paillier homomorphic cryptosystems \cite{paillier1999public,Zhu-PrivacyPreservingTimeSeriesData-EDBT-2014}.
It should be finally noted that, while conventional image based biometric systems employ non-generative comparators, operating either on binary or nonnegative integers \cite{GomezBarrero-MBTPwithHE-PR-2017,GomezBarrero-VariableLengthBTPHE-Access-2017,penn2014customisation,barniprivacy}, the generative comparators used in speaker recognition applications make assumptions on underlying distributions, such as normal distribution \cite{Dehak-FrontEndFactorAnalysis-ivector-TASLP-2011,Kenny-JFA-Theory-CRIM-2005}, consequently operating on normal distributed float values.

In the following, we make HE available to speaker recognition, targeting data privacy for subjects and vendors.
Secs.~\ref{sec:related_work}, \ref{sec:HE_cryptosystems}, \ref{sec:speaker_recognition} depict related work on homomorphic cryptosystems and speaker recognition. 
Sec.~\ref{sec:proposed} proposes two architectures for HE protected 2Cov comparators.
A proof\-/of\-/concept study is discussed in Sec.~\ref{sec:experiments} with conclusions drawn in Sec.~\ref{sec:conclusion}.

\section{Related Work}
\label{sec:related_work}

In order to apply standardized biometric template protection schemes, binarization can be employed \cite{ISO-IEC-24745-TemplateProtection-101129}. Related work on the binarization of traditional speaker recognition systems utilizing universal background models (UBMs) targeting the GMM\,--\,UBM approach can be found in \cite{AngueraBonastre-SpeakerBinaryKey-Interspeech-2010,Bonastre-DiscriminantBinaryData-ICASSP-2011,Hernandez-SpeakerBinaryRepresentation-ProgressPatternRecognition-2012}. In addition, in our earlier work \cite{Paulini-MultiBitAllocation-Odyssey-2016}, we proposed a biometric template protection scheme for speaker recognition, based on binarized Gaussian mixture model (GMM) supervectors.

However, due to the binarization process, the biometric performance usually declines, and calibration properties are lost.
Contrary to performance\-/lossy template protection approaches as biometric cryptosystems and cancelable biometrics~\cite{Rathgeb-BTP-Survey-EURASIP-2011}, HE completely preserves biometric accuracy. 
Therefore, we investigate on Paillier HE schemes, which are already introduced to other biometric modalities, such as signature~\cite{gomez2016implementation}, iris~\cite{penn2014customisation}, and fingerprint~\cite{barniprivacy} recognition, considering Hamming distances (XOR operator), dynamic time warping (DTW), the Euclidean distance, and the cosine similarity. We thus focus on homomorphic cryptosystems for the remainder of the article.

In \cite{Barni-BTP-SPM-2015} and \cite{Bringer-BTP-Survey-SPM-2013}, the authors provide an overview of several biometric template protection schemes based on homomorphic encryption and garbled circuits. Barni et al.~\cite{barniprivacy} present a way to protect fixed-length fingercodes~\cite{Jain-Fingercodes-IEEE-1997} using homomorphic encryption. This system was modified in \cite{Bianchi-BTP-Fingercode-BIOMS-2010} to accelerate the process by reducing the size of the fingercode. However, a reduction of information also leads to a degradation of biometric recognition performance. 
Ye et al. present an anonymous biometric access control (ABAC) system~\cite{Ye-BTP-HE-EURASIP-2009} for iris recognition. Their system setup verifies only whether a subject is enrolled without revealing the identity and thus grants anonymity towards the subjects. Another ABAC protocol is proposed in \cite{Luo-BTP-HE-ICME-2009} by Luo \mbox{et al.} and a secure similarity search algorithm is presented for anonymous authentication.
Combining homomorphic encryption with garbled circuits, Blanton and Gasti~\cite{Blanton-BTP-IrisFingerprint-ESORICS-2011} implement secure protocols for iris and fingerprint recognition.

Among the existing cryptosystems in the literature, encryption algorithms based on lattices are assumed to be post\-/quantum secure~\cite[conjecture~2]{quantum}, which is a convenient property for a public key encryption scheme. 
Using ideal lattices in a somewhat homomorphic encryption (SHE) scheme, Yasuda \etal~\cite{Yasuda-BTP-HE-ARES-2013} compute the Hamming distance of encrypted templates in an efficient way by using a packing method before the encryption.
By using binary feature vectors with a constant size of $2\,048\bits$ for every biometric data, again Yasuda \etal~\cite{Yasuda-BTP-HE-SCN-2015} present a new packing method in a SHE scheme for biometric authentication based on a special version of the ring learning with errors assumption.
Another privacy\-/preserving biometric authentication approach~\cite{Patsakis-BTP-HE-DPM-2015} splits a $2\,048\bits$ iris code into $64$ blocks with $32\bits$ each and encrypts these blocks using \mbox{n\-/th} degree truncated polynomial ring (NTRU). As in the aforementioned works, scores are computed in the encrypted domain without disclosure of biometric information.

\section{Homomorphic Cryptosystems}
\label{sec:HE_cryptosystems}
Homomorphic encryption~\cite{Rivest-PrivacyHomomorphisms-SecureComputation-1978,FontaineGaland-Survey-HomomorphicEncryption-JIS-2007,Hoffstein-IntroductionCryptography-2008} has the property that computations on the ciphertext are equivalent to those carried out on the plaintext.
In particular, homomorphisms are functions which preserve algebraic structures of groups \cite{Hungerford-Algebra-Springer-1974}.
The function $f: G \rightarrow H$ is a homomorphism for groups $(G, \diamond), (H, \smallsquare)$ with sets $G, H$ and operators $\diamond, \smallsquare$ if:
\begin{equation}
f(g \diamond g') = f(g) \smallsquare f(g') \qquad \forall g,g' \in G.
\end{equation}

Public\-/key cryptosystems $(K, M, C, \enc, \dec)$ with sets of keys $K$, plaintexts $M$, ciphertexts $C$, and functions representing encryption $\enc$ and decryption $\dec$ are homomorphic if:
\begin{align}
\forall m_1,m_2 \in M, &\forall {\emph{pk}} \in K: \nonumber\\
\enc_{\emph{pk}}(m_1) \smallsquare \enc_{\emph{pk}}(m_2) &= \enc_{\emph{pk}}(m_1 \diamond m_2),
\end{align}
where the public key ${\emph{pk}}$ is used for encryption and the secret key ${\emph{sk}}$ for the decryption functions, respectively:
\begin{align}
\enc_{\emph{pk}}: M \rightarrow C,\nonumber\\
\dec_{\emph{sk}}: C \rightarrow M.
\end{align}

\subsection{Paillier HE Scheme}
Motivated by asymmetric Paillier cryptosystems \cite{paillier1999public,Zhu-PrivacyPreservingTimeSeriesData-EDBT-2014}, HE has been made available to biometric template protection \cite{gomez2016implementation,GomezBarrero-MBTPwithHE-PR-2017,GomezBarrero-VariableLengthBTPHE-Access-2017}.
Paillier cryptosystems are homomorphic probabilistic encryption schemes based on the decisional composite residuosity assumption (DCRA): for integers $n,z$ it is hard to decide, whether $z$ is an $n$-residue modulo $n^2$. Due to this assumption, the Paillier cryptosystem is secure against \emph{honest but curious users} conducting chosen ciphertext attacks \cite{paillier1999public,bellare1998relations,paillier1999efficient}.

In the Paillier cryptosystem, the public key ${\emph{pk}} = (n,g)$ is defined by $n = p\,q$ and $g\in\Z^*_{n^2}$, where $p,q$ are two large prime numbers, such that $\gcd(p\,q,(p-1)\,(q-1)) = 1$, and with $\Z^*_{n^2}$ as the set of module $n^2$ integers having a modular multiplicative inverse.
The modular multiplicative inverse $\overline{\varrho}$ to $\varrho$ is required with $\gcd(\varrho, \overline{\varrho}) = 1$: $\varrho\,\overline{\varrho} \equiv 1 \pmod{n^2}$.
Based on $p,q$, the secret key ${\emph{sk}} = (\lambda, \mu)$ is defined by $\lambda = \lcm(p-1,q-1)$ and $\mu = \overline{\varrho} \mod n$ with $\varrho = L(g^{\lambda} \mod n^2)$ and $L(x) = \frac{x-1}{n}$.

During encryption $c = \enc_{\emph{pk}}(m,s) \in \Z^*_{n^2}$ of a message $m\in\Z_n$ with public key ${\emph{pk}}$, a random number $s \in \Z^*_n$ provides the probabilistic nature of the cryptosystem, \ie $\enc_{\emph{pk}}(m, s_1) \neq \enc_{\emph{pk}}(m, s_2)$ for two different $s_1,s_2 \in \Z^*_n$:
\begin{equation}
c = \enc_{\emph{pk}}(m,s) = g^m\,s^n \mod{n^2},
\end{equation}
which is abbreviated in the following as $\enc_{\emph{pk}}(m)$. 

Ciphertexts are decrypted as:
\begin{equation}
m = \dec_{\emph{sk}}(c) = L\left(c^{\lambda} \mod{n^2}\right)\,\mu \mod{n}.
\end{equation}

Similarly to \cite{gomez2016implementation,GomezBarrero-MBTPwithHE-PR-2017,GomezBarrero-VariableLengthBTPHE-Access-2017,Zhu-PrivacyPreservingTimeSeriesData-EDBT-2014}, we utilize the additive homomorphic properties of the Paillier cryptosystem regarding plaintexts $m_1,m_2$ and corresponding ciphertexts $c_1,c_2$:
\begin{align}
\dec_{\emph{sk}}\left(c_1\,c_2\right) &= m_1 + m_2 \mod{n},\nonumber\\
\dec_{\emph{sk}}\left({c_1}^l\right) &= m_1\,l \mod{n}, \text{ with a constant $l$}.
\label{eq:HE:properties}
\end{align}
In other words, whereas the decrypted product of two ciphertexts is equivalent to the sum of two plaintexts, the corresponding exponentiation of a ciphertext results in the product of the corresponding plaintext and constant as exponent.

\subsection{Homomorphic Template Protection}
Targeting biometric template protection, data privacy friendly comparison schemes are sought, in which only encrypted references, \ie no plaintexts, are stored in databases.
As such, the Euclidean and cosine similarity comparison scores $S_{\emph{Euc}}, S_{\emph{cos}}$ between two $F$-dimensional vectors $\bm{X} = \{x_1, \dots, x_F\}, \bm{Y} = \{y_1, \dots, y_F\}$ are computationally derived as \cite{gomez2016implementation,GomezBarrero-MBTPwithHE-PR-2017,GomezBarrero-VariableLengthBTPHE-Access-2017}:
\begin{align}
S_{\emph{Euc}}\left(\bm{X}, \bm{Y}\right) &= \sum\limits_{f=1}^F x_f^2 + \sum\limits_{f=1}^F y_f^2 - 2\,\sum\limits_{f=1}^F x_f\,y_f,
\end{align}
and the corresponding encrypted score $\enc_{\emph{pk}}\left(S_{\emph{Euc}}\left(\bm{X}, \bm{Y}\right)\right)$:
\begin{align}
\enc_{\emph{pk}}\left(S_{\emph{Euc}}\left(\bm{X}, \bm{Y}\right)\right) &= \nonumber\\
 \enc_{\emph{pk}}\left(\sum\limits_{f=1}^F x_f^2\right)\,\enc_{\emph{pk}}\left(\sum\limits_{f=1}^F y_f^2\right)&\,\prod\limits_{f=1}^F\enc_{\emph{pk}}\left(y_f\right)^{-2\,x_f},
\end{align}
where the protected reference $\bm{Y}_{\emph{Euc}}^{\enc_{\emph{pk}}}$ is defined as:
\begin{equation}
\bm{Y}_{\emph{Euc}}^{\enc_{\emph{pk}}} = \Bigg(\enc_{\emph{pk}}\left(\sum\limits_{f=1}^F y_f^2\right), \left(\enc_{\emph{pk}}\left(y_f\right)\right)_{f=1}^F\Bigg).
\end{equation}
On the other hand, the cosine comparison is derived as \cite{gomez2016implementation,GomezBarrero-MBTPwithHE-PR-2017}:
\begin{align}
S_{\emph{cos}}\left(\bm{X}, \bm{Y}\right) &= \frac{\bm{X'}\,\bm{Y}}{||\bm{X}||\,||\bm{Y}||} = \sum\limits_{f=1}^F\frac{x_f}{||\bm{X}||}\frac{y_f}{||\bm{Y}||},\nonumber\\
\enc_{\emph{pk}}\left(S_{\emph{cos}}\left(\bm{X}, \bm{Y}\right)\right) &= \prod\limits_{f=1}^F\enc_{\emph{pk}}\left(\frac{y_f}{||\bm{Y}||}\right)^{\frac{x_f}{||\bm{X}||}},
\end{align}
where the protected reference $\bm{Y}_{\emph{cos}}^{\enc_{\emph{pk}}}$ is defined for length\-/normalized features as:
\begin{equation}
\bm{Y}_{\emph{cos}}^{\enc_{\emph{pk}}} = \left(\left(\enc_{\emph{pk}}\left(y_f\right)\right)_{f=1}^F\right) = \enc_{\emph{pk}}(\bm{Y}).
\end{equation}
In \cite{gomez2016implementation,GomezBarrero-MBTPwithHE-PR-2017}, solely positive integers are considered.
Accommodating a broader range of only positive float values, a $10^{12}$ scaling factor is employed.
Accounting for negative float values, this study relies on an alternative float representation.

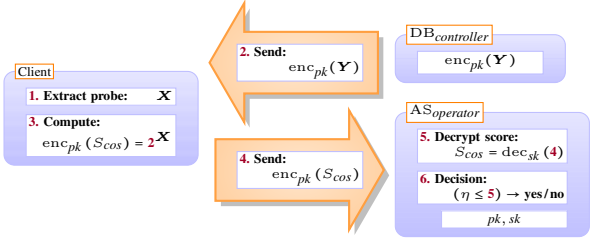
\begin{figure}[htb]
    \centering
    \begin{tikzpicture}[font=\tiny, node distance=1em]
    \node [module step] 
        (step1) {\stepno{1.} \textbf{Extract probe:} \hfill\matharchitecturenode{$\bm{X}$}};
    \node [below=1.25em of step1.south west,module step] 
        (step3) {\stepno{3.} \textbf{Compute:} \\\hfill\matharchitecturenode{$\enc_{\emph{pk}}\left(S_{\emph{cos}}\right) = \stepno{2}^{\bm{X}}$}};
    \pic {module={Client}{step1}{step3}};
    \coordinate (ClientEast) at ([yshift=2.5em]$(step1.north east)!0.5!(step3.south east)$);
    \coordinate (DB1) at ([xshift=9em]ClientEast);
    \node [right=of DB1,module data]
        (data1) {\matharchitecturenode{$\enc_{\emph{pk}}(\bm{Y})$}};
    \pic {module={$\mathrm{DB}_{\emph{controller}}$}{data1}{data1}};
    \coordinate (Comm1) at ([xshift=-2em]$(ClientEast)!0.5!(DB1)$);
    \node [at=(Comm1),channel] 
        (step2) {\stepno{2.} \textbf{Send:} \\\hfill\matharchitecturenode{$\enc_{\emph{pk}}(\bm{Y})$}};
    \pic {channel left={Comm1}{2.75em}{0em}{7em}{5em}{.5em}{120}};
    \coordinate (Comm2) at ([yshift=-4.5em]Comm1);
    \node [at=(Comm2),channel,rotate=0,xshift=0em] 
    (step4) {\stepno{4.} \textbf{Send:} \\\hfill\matharchitecturenode{$\enc_{\emph{pk}}\left(S_{\emph{cos}}\right)$}};
    \pic {channel right={Comm2}{2.25em}{0em}{7em}{5em}{.5em}{120}};
     Application Server
    \coordinate (AS1) at ([yshift=1em]DB1|-Comm2);
    \node [right=of AS1,module step] 
        (step5) {\stepno{5.} \textbf{Decrypt score:} \\ \hfill\matharchitecturenode{$S_{\emph{cos}}=\dec_{\emph{sk}}\left(\stepno{4}\right)$}};
    \node [below=of step5.south west,module step] 
        (step6) {\stepno{6.} \textbf{Decision:} \\ \hfill\matharchitecturenode{$(\eta \leq \stepno{5}) \rightarrow$ \textbf{yes\,/\,no}}};
    \node [below=of step6.south east,module data,anchor=east,yshift=0.375em]
        (data2) {\matharchitecturenode{${\emph{pk}}, {\emph{sk}}$}};
    \pic {module={$\mathrm{AS}_{\emph{operator}}$}{step5}{data2}};
\end{tikzpicture}
    \caption{Architecture of homomorphic encrypted cosine similarity comparison for length\-/normalized features, \cf~\cite{gomez2016implementation}, with client, servers (blue) and communication channels (orange).}
    \label{fig:cos_arch}
\end{figure}

Fig.~\ref{fig:cos_arch} illustrates a distributed client\,--\,server architecture employing HE with a cosine comparison: a client $C$ extracts the probe feature vector $\bm{X}$ and requests the encrypted reference feature vector $enc_{\emph{pk}}(\bm{Y})$ from the database $\mathrm{DB}_{\emph{controller}}$.
Then, scores are calculated on the client, and sent to the authentication server $\mathrm{AS}_{\emph{operator}}$, which holds the key pair $({\emph{pk}}, {\emph{sk}})$. Based on pre\-/defined threshold, $\mathrm{AS}_{\emph{operator}}$ outputs the decision $D$ of whether the decrypted score $S_{\emph{cos}}$ is greater or equal to a threshold~$\eta$, or not.
Ideally, $\mathrm{DB}_{\emph{controller}}$ is in the domain of an independent data controller, restricting access to operators among others.
Tab.~\ref{tab:complexityRelated} provides an overview to the complexity of the encrypted Euclidean and cosine comparison, where numbers diverge from \cite{gomez2016implementation} as in the signature recognition scenario, five reference templates are encrypted rather than \eg, an averaged template model.
As references are encrypted during enrolment, cosine\-/based biometric comparisons require no additional encryptions, whereas in Euclidean\-/based comparisons, the probe templates need to be encrypted.

\begin{table}[htb]
    \caption{Complexity analysis for the Euclidean and cosine comparators during verification, \cf \cite{gomez2016implementation}, assuming $F=250$ dimensional features, the size of an encrypted feature $c=0.5\KiB$, and the plain feature size $p=64\bits$.}
    \label{tab:complexityRelated}
    \scriptsize
    \centering
    \smallskip
    \begin{tabular}{lcc} 
        \toprule
        & Euclidean & Cosine \\
        \midrule
        N$^o$ encryptions & $F$ & $0$\\
        N$^o$ decryptions & $1$ & $1$\\
        \midrule
        N$^o$ additions & $F-1$ & $0$ \\
        N$^o$ products & $2\,F+4$ & $F - 1$ \\
        N$^o$ exponentiations & $2\,F$ & $F$ \\
        \midrule
        Plain template size & $p\,F$ & $p\,F$ \\
        & $\approx 2.0\KiB$  & $\approx 2.0\KiB$ \\
        Protected template size & $c\,(F+1)$ & $c\,F$ \\
        & $= 125.5\KiB$ & $= 125.0\KiB$ \\
        \midrule
        Channels: amount of & $c\,(F+2)$ & $c\,(F + 1)$ \\
        protected data exchanged & $= 126.0\KiB$ & $= 125.5\KiB$ \\
        \bottomrule
    \end{tabular}
\end{table}

\section{Speaker Recognition: 2Cov Comparator}
\label{sec:speaker_recognition}
Recent speaker recognition approaches rely on intermediate\-/sized vectors (i-vectors), representing the characteristic speaker offset from an UBM, which models the distribution of acoustic features, such as Mel-Frequency Cepstral Coefficients (MFCCs) \cite{Reynolds-GMMs-ConversationalSpeech-2000}.
UBM components' mean vectors are concatenated to a \emph{supervector} $\bm{\mu_{\text{UBM}}}$.
Seeking non\-/sparse features, speaker supervectors \(\bm{s}\) are decomposed by a total variability matrix $\mathbf{T}$ into a lower\-/dimensional and higher\-/discriminant \mbox{i\-/vectors} $\bm{i}$ as an offset to the UBM supervector $\bm{\mu_{\text{UBM}}}$:
\begin{align}
    \label{eq:ivec}
    \bm{s} = \bm{\mu_{\text{UBM}}} + \bm{T}\,\bm{i}.
\end{align}
The total variability matrix is trained on a development set using an expectation maximization algorithm \cite{Dehak-FrontEndFactorAnalysis-ivector-TASLP-2011,Kenny-JFA-Theory-CRIM-2005}. 
Then, \mbox{i-vectors} are projected onto a unit\-/spherical space by whitening transform and length-normalization \cite{GarciaRomeroEpsyWison-ivector-lengthnormalization-Interspeech-2011,Bousquet-BenefitsDifferentSteps-ivectors-CIARP-2013}.

State-of-the-art \mbox{i-vector} comparators belong to the PLDA family \cite{Cumani-GenerativePairwiseModels-Odyssey-2014, Bousquet-BenefitsDifferentSteps-ivectors-CIARP-2013}. 
PLDA comparators conduct a likelihood ratio scoring comparing the probabilities of the hypotheses that reference and probe \mbox{i-vectors} $\bm{X}, \bm{Y}$ stem from (a) the same source or (b) different sources.
Therefore, within and between speaker variabilities are examined in a latent feature subspace.
In this work, emphasis is put on the 2Cov approach \cite{Cumani-PairwiseDiscriminativeVerification-TASLP-2013,Cumani-GenerativePairwiseModels-Odyssey-2014}, the full\-/subspace Gaussian PLDA.
Notably, the 2Cov comparator can also be related to pairwise support vector machines \cite{Cumani-PairwiseDiscriminativeVerification-TASLP-2013,Cumani-GenerativePairwiseModels-Odyssey-2014}.
For the sake of tractability, this study focuses on the generative 2Cov model.
Also, \mbox{i-vectors} are solely considered as point estimates, assuming ideal precision during feature extraction.
The closed-form solution to the 2Cov scoring is denoted regarding within and between covariances $\bm{W^{-1}}, \bm{B^{-1}}$ with mean~$\bm{\mu}$~\cite{Cumani-PairwiseDiscriminativeVerification-TASLP-2013}:
\begin{align}
    S_{\emph{2Cov}}\left(\bm{X}, \bm{Y}\right) &= 
    \bm{X'}\,\bm{\Lambda}\,\bm{Y} + \bm{Y'}\,\bm{\Lambda}\,\bm{X}+ \bm{X'}\,\bm{\Gamma}\,\bm{X} + \nonumber\\
    &\hphantom{{} = {}} \bm{Y'}\,\bm{\Gamma}\,\bm{Y} + \bm{c'}\,\left(\bm{X} + \bm{Y}\right) + k,\nonumber\\
    \bm{\Lambda} &= \frac{1}{2}\bm{W'}\,\bm{\tilde{\Lambda}}\,\bm{W},
    \qquad
    \bm{\Gamma} = \frac{1}{2}\bm{W'}\,\left(\bm{\tilde{\Lambda}}-\bm{\tilde{\Gamma}}\right)\,\bm{W},\nonumber\\
    \bm{c} &= \bm{W'}\,\left(\bm{\tilde{\Lambda}}-\bm{\tilde{\Gamma}}\right)\,\bm{B}\,\bm{\mu},
    \nonumber\\
    k &= \tilde{k} + \frac{1}{2}\left(\left(\bm{B}\,\bm{\mu}\right)'\,\left(\bm{\tilde{\Lambda}} - 2\,\bm{\tilde{\Gamma}}\right)\,\bm{B}\,\bm{\mu}\right),\nonumber\\
    \bm{\tilde{\Lambda}} &= \left(\bm{B} + 2\,\bm{W}\right)^{-1},
    \quad
    \bm{\tilde{\Gamma}} = \left(\bm{B}+\bm{W}\right)^{-1},\nonumber\\
    \tilde{k}&= 2\,\log|\bm{\tilde{\Gamma}}| - \log|\bm{\tilde{\Lambda}}| - \log|\bm{B}| + \bm{\mu'}\,\bm{B}\,\bm{\mu}.
    \label{eq:HE:2CovCumani}
\end{align}

\section{Proposed Architecture}
\label{sec:proposed}

In the following, two discriminative HE schemes are proposed.
The first puts emphasis on HE for i\-/vectors during 2Cov comparison, seeking data privacy for end\-/users, whereas the second scheme focuses on the encryption of i\-/vectors as well as 2Cov model parameters, targeting data protection for subjects and vendors.
An auxiliary float representation is implemented, encoding float values as nonnegative integers for the purpose of providing Paillier properties, \cf Eq.~\eqref{eq:HE:properties}.

\subsection{Auxiliary Float Representation: nonnegative Integers}
For the purpose of representing float values of i\-/vectors as nonnegative integer values, \ie seeking conformance to Paillier cryptosystems, the integer encoding scheme standardized in IEEE~754 is employed \cite{ieee2008754}.
Floats are encoded in terms of a sign $S$, a mantissa $M$ times a base $B=16$ raised to an exponent $E$.
Nonnegative integers are derived by seeking congruent positive representations in modulo $n^2$, \ie regarding the public key domain.
Accounting for negative values \cite{Thorne-PythonPaillier-2017}, the plaintext integer domain is divided into four intervals: $[0, \frac{n}{3})$ for positive float representations, $[\frac{2\,n}{3}, n)$ for negative float representations, and $[\frac{n}{3}, \frac{2\,n}{3})$ as well as $[n, \infty)$ for the purpose of detecting overflows resulting from previous Paillier HE operations.
Targeting Paillier HE, same exponents of $m_1, m_2$ are required, hence the mantissa is encrypted as a nonnegative integer representation.
The plaintext exponent of the depending mantissa encoding is kept auxiliary.
Security is satisfied due to the DCRA employing randomized mantissa obfuscation during encryption.
In Paillier addition, encrypted mantissae are scaled for equivalent addend exponents.
In Paillier multiplication, modular exponentiation of $c = \enc_{\emph{pk}}(M, s)$ is conducted, during which mantissae are kept rather small by iterative multiplications than by right\-/away exponentiation.

\subsection{Data Privacy: Protecting Subjects}
For the sake of tractability, we assume a zero mean, causing $\bm{c} = \bm{0}$, and neglect the normalization term, \ie $k=0$, such that the following scheme solely holds for discriminative 2Cov, however calibrated scores can be easily achieved by adding the $k$ term after score decryption:
\begin{align}
S_{\emph{2Cov}}\left(\bm{X}, \bm{Y}\right) &= 
\bm{X'}\,\bm{\Lambda}\,\bm{Y} + \bm{Y'}\,\bm{\Lambda}\,\bm{X}+ \bm{X'}\,\bm{\Gamma}\,\bm{X} + \bm{Y'}\,\bm{\Gamma}\,\bm{Y},\nonumber\\
&= 
\left(\bm{X'}\,\bm{\Lambda}\right)\,\bm{Y} + 
\bm{Y'}\,\left(\bm{\Lambda}\,\bm{X}\right)+ \nonumber\\
&\hphantom{{} = {}}
\bm{X'}\,\bm{\Gamma}\,\bm{X} + 
\bm{Y'}\,\bm{\Gamma}\,\bm{Y}.
\end{align}

For the discriminative 2Cov, HE is employed motivated by the (symmetric) dot product for vector multiplication:
\begin{align}
\enc_{\emph{pk}}(\bm{Y})^{\bm{X}} 
&= \prod\limits_{f=1}^F \enc_{\emph{pk}}(y_f)^{x_f} 
= \enc_{\emph{pk}}(\bm{X'}\,\bm{Y}) \nonumber\\
= \enc_{\emph{pk}}(\bm{Y'}\,\bm{X})
&= \prod\limits_{f=1}^F \enc_{\emph{pk}}(x_f)^{y_f} 
= \enc_{\emph{pk}}(\bm{X})^{\bm{Y}},\nonumber\\
%
\enc_{\emph{pk}}\left(S_{\emph{2Cov}}\left(\bm{X}, \bm{Y}\right)\right) &= \enc_{\emph{pk}}(\bm{Y})^{(\bm{X'}\,\bm{\Lambda})} \,
\enc_{\emph{pk}}(\bm{Y})^{(\bm{\Lambda}\,\bm{X})}
\,\nonumber\\
&\hphantom{{} = {}} \enc_{\emph{pk}}(\bm{X'\,\bm{\Gamma}\,\bm{X}})\,
\enc_{\emph{pk}}(\bm{Y'\,\bm{\Gamma}\,\bm{Y}})\,
,\nonumber\\
\enc_{\emph{pk}}(\bm{Y}) &= \left(\enc_{\emph{pk}}(y_f)\right)_{f=1}^F,
\label{eq:HE:2CovEncrypted}
\end{align}
with auxiliary vectors are denoted as $\left(\bm{X'}\,\bm{\Lambda}\right), \left(\bm{\Lambda}\,\bm{X}\right)$, and the protected reference $\bm{Y}^{\enc_{\emph{pk}}}_{2Cov} = \left(\enc_{\emph{pk}}(\bm{Y}), \enc_{\emph{pk}}(\bm{Y'\,\bm{\Gamma}\,\bm{Y}})\right).$

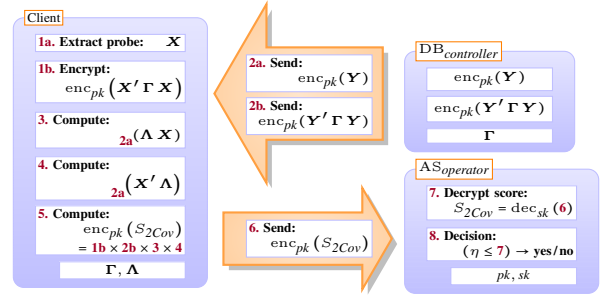
\begin{figure}[htb]
    \centering
    \begin{tikzpicture}[font=\tiny, node distance=1em]
    \node [module step] 
        (step1a) {\stepno{1a.} \textbf{Extract probe:} \hfill\matharchitecturenode{$\bm{X}$}};
    \node [below=1.25em of step1a.south west,module step] 
        (step1b) {\stepno{1b.} \textbf{Encrypt:}\\\hfill \matharchitecturenode{$\enc_{\emph{pk}}\left(\bm{X'}\,\bm{\Gamma}\,\bm{X}\right)$}};
    \node [below=1.125em of step1b.south west,module step] 
        (step3) {\stepno{3.} \textbf{Compute:}\\\hfill \matharchitecturenode{$\stepno{2a}^{\left(\bm{\Lambda}\,\bm{X}\right)}$}};
    \node [below=1.25em of step3.south west,module step] 
        (step4) {\stepno{4.} \textbf{Compute:}\\\hfill \matharchitecturenode{$\stepno{2a}^{\left(\bm{X'}\,\bm{\Lambda}\right)}$}};
    \node [below=1.25em of step4.south west,module step] 
        (step5) {\stepno{5.} \textbf{Compute:}\\ \hfill{\matharchitecturenode{$\enc_{\emph{pk}}\left(S_{\emph{2Cov}}\right)$}\\\hfill\matharchitecturenode{$={}\stepno{1b}\times\stepno{2b}\times\stepno{3}\times\stepno{4}$}}};
    \node [below=of step5.south east,module data,anchor=east,yshift=0.375em]
        (dataClient) {\matharchitecturenode{$\bm{\Gamma}, \bm{\Lambda}$}};
    \pic {module={Client}{step1}{dataClient}};
    \coordinate (ClientEast) at ([yshift=1em]$(step1.north east)!0.5!(step3.south east)$);
    \coordinate (ClientEastSouth) at ([yshift=2em]$(step4.north east)!0.5!(dataClient.south east)$);
    \coordinate (DB1) at ([xshift=9em,yshift=-0.5em]ClientEast);
    \node [right=of DB1,module data]
        (data1) {\matharchitecturenode{$\enc_{\emph{pk}}(\bm{Y})$}};
    \node [below=1.25em of data1.east,module data,anchor=east,yshift=0.0em]
        (data2) {\matharchitecturenode{$\enc_{\emph{pk}}(\bm{Y'}\,\bm{\Gamma}\,\bm{Y})$}};
    \node [below=1.25em of data2.east,module data,anchor=east,yshift=0.125em]
        (data3) {\matharchitecturenode{$\bm{\Gamma}$}};
    \pic {module={$\mathrm{DB}_{\emph{controller}}$}{data1}{data3}};
    \coordinate (Comm1) at ([xshift=-2em,yshift=0em]$(ClientEast)!0.5!(DB1)$);
    \node [at=(Comm1),channel] 
        (step2a) {\stepno{2a.} \textbf{Send:} \\\hfill\matharchitecturenode{$\enc_{\emph{pk}}(\bm{Y})$}};
    \node [below=of step2a.south west,channel] 
        (step2b) {\stepno{2b.} \textbf{Send:} \\\hfill\matharchitecturenode{$\enc_{\emph{pk}}(\bm{Y'}\,\bm{\Gamma}\,\bm{Y})$}};
    \pic {channel left={Comm1}{2.75em}{-.85em}{7em}{7.5em}{.5em}{120}};
    \coordinate (AS1) at ([xshift=10em,yshift=-0.375em]ClientEastSouth);
    \node [below=of AS1,module step] 
        (step7) {\stepno{7.} \textbf{Decrypt score:} \\ \hfill\matharchitecturenode{$S_{\emph{2Cov}}= \dec_{\emph{sk}}\left(\stepno{6}\right)$}};
    \node [below=of step7.south west,module step] 
        (step8) {\stepno{8.} \textbf{Decision:} \\ \hfill\matharchitecturenode{$(\eta \leq \stepno{7}) \rightarrow$ \textbf{yes\,/\,no}}};
    \node [below=of step8.south east,module data,anchor=east,yshift=0.375em]
        (dataASoperator) {\matharchitecturenode{${\emph{pk}}, {\emph{sk}}$}};
    \pic {module={$\mathrm{AS}_{\emph{operator}}$}{step7}{dataASoperator}};
    \coordinate (Comm2) at ([yshift=-2.5em]Comm1|-AS1);
    \node [at=(Comm2),channel,rotate=0,xshift=0em] 
        (step6) {\stepno{6.} \textbf{Send:} \\\hfill\matharchitecturenode{$\enc_{\emph{pk}}\left(S_{\emph{2Cov}}\right)$}};
    \pic {channel right={Comm2}{2.25em}{0em}{7em}{4.25em}{.5em}{120}};
\end{tikzpicture}
    \caption{Architecture of homomorphic encrypted 2Cov comparison solely protecting subject data, with client, servers (blue) and communication channels (orange).}
    \label{fig:HE:2Cov_schema_subject}
\end{figure}

Fig.~\ref{fig:HE:2Cov_schema_subject} illustrates the proposed HE architecture for a distributed system.
Similarly to the cosine comparison HE approach, the scores are computed in the encrypted domain on the client, and decrypted on the authentication server.
Thereby, the 2Cov score is computed in four parts.

\subsection{Data Privacy: Protecting Subjects and Vendors}
Contrary to established biometric HE approaches employing non\-/generative comparators, generative comparators require trained hyper\-/parameters \eg, between and within covariance matrices in terms of the 2Cov comparator.
For the purpose of protecting both subject and vendor data, two key sets are employed $(pk1, sk1), (pk2, sk2)$.
Utilizing the Frobenius inner product\footnote{The inner Frobenius product denotes $\bm{x'}\,\bm{A}\,\bm{y} = \langle\bm{A}, \bm{x}\,\bm{y'}\rangle = \mathrm{vec}(\bm{A})'\,\mathrm{vec}(\bm{x}\,\bm{y'})$, where $\mathrm{vec}(\cdot)$ denotes the operator stacking matrices into a vector and $\langle\bm{A}, \bm{B}\rangle$ is the dot product between matrices, \cf~\cite{Cumani-PairwiseDiscriminativeVerification-TASLP-2013}.}, Eq.~\eqref{eq:HE:2CovCumani} can be reformulated \cite{Cumani-PairwiseDiscriminativeVerification-TASLP-2013}:
\begin{align}
S_{\emph{2Cov}}\left(\bm{X}, \bm{Y}\right) &= 
\langle\bm{\Lambda}, \bm{X}\,\bm{Y'} + \bm{Y}\,\bm{X'}\rangle+ \langle\bm{\Gamma}, \bm{X}\,\bm{X'} + \bm{Y}\,\bm{Y'}\rangle +\nonumber\\
&\hphantom{{} = {}}
\bm{c'}\,\left(\bm{X} + \bm{Y}\right) + k,\nonumber\\
&= \bm{w_{\Lambda}'}\,\varphi_{\bm{\Lambda}}(\bm{X}, \bm{Y}) + 
\bm{w_{\Gamma}'}\,\varphi_{\bm{\Gamma}}(\bm{X}, \bm{Y}) + \nonumber\\
&\hphantom{{} = {}} \bm{w_{c}'}\,\varphi_{\bm{c}}(\bm{X}, \bm{Y}) + 
\bm{w'}_{k}\,\varphi_{k}(\bm{X}, \bm{Y}),\nonumber\\
&= \bm{w'}\,\varphi(\bm{X}, \bm{Y}), \textit{ with:}\nonumber\\
\varphi(\bm{X}, \bm{Y}) &= 
\begin{bmatrix}
\mathrm{vec}(\bm{X}\,\bm{Y'} + \bm{Y}\,\bm{X'})\\
\mathrm{vec}(\bm{X}\,\bm{X'} + \bm{Y}\,\bm{Y'})\\
\bm{X} + \bm{Y}\\
1
\end{bmatrix}
= 
\begin{bmatrix}
\varphi_{\bm{\Lambda}}(\bm{X}, \bm{Y})\\
\varphi_{\bm{\Gamma}}(\bm{X}, \bm{Y})\\
\varphi_{\bm{c}}(\bm{X}, \bm{Y})\\
\varphi_{k}(\bm{X}, \bm{Y})
\end{bmatrix},\nonumber\\
\bm{w} &= 
\begin{bmatrix}
\mathrm{vec}(\bm{\Lambda})\\
\mathrm{vec}(\bm{\Gamma})\\
\bm{c}\\
k
\end{bmatrix}
=
\begin{bmatrix}
\bm{w_{\Lambda}}\\
\bm{w_{\Gamma}}\\
\bm{w_{c}}\\
\bm{w}_{k}
\end{bmatrix}.
\label{eq:HE:2CovFrobenius}
\end{align}

For the simplified 2Cov comparator, a mutual HE scheme sustaining data privacy for subjects and vendors can be employed by extending the inner product of vectors to the Frobenius inner product of matrices $\bm{A}, \bm{B}$, which can be reformulated via the $\mathrm{vec}(\cdot)$ operator as the inner product of (column-stacked) vectors, such that the dot product can be employed as well with a public key ${\emph{pk}}$:
\begin{equation}
\enc_{\emph{pk}}\left(\bm{A}\right)^{\langle\rangle(\bm{B})} = \enc_{\emph{pk}}\left(\mathrm{vec}\left(\bm{A}\right)\right)^{\mathrm{vec}\left(\bm{B}\right)},
\end{equation}
where the encryption of a matrix $\bm{A}$ is denoted as:
\begin{equation}
\enc_{\emph{pk}}\left(\bm{A}\right) = \left(\left(\enc_{\emph{pk}}(a_{i,j})\right)_{i=1}^F\right)_{j=1}^F.
\end{equation}

In the simplified 2Cov comparator, the encrypted vendor and operator communication takes the form:
\begin{align}
S_{\emph{2Cov}}\left(\bm{X}, \bm{Y}\right) &= \bm{w_{\Lambda}'}\,\varphi_{\bm{\Lambda}}(\bm{X}, \bm{Y}) + 
\bm{w_{\Gamma}'}\,\varphi_{\bm{\Gamma}}(\bm{X}, \bm{Y}),\nonumber\\
\enc_{\emph{pk2}}\left(S_{\emph{2Cov}}\left(\bm{X}, \bm{Y}\right)\right) &= \enc_{\emph{pk2}}(\bm{\Lambda})^{\langle\rangle(\bm{c_1})}\!\enc_{\emph{pk2}}(\bm{\Gamma})^{\langle\rangle\left(\bm{c_2} + \bm{c_3}\right)}, \nonumber\\
\textit{ with: }\bm{c_1} = \bm{X}\,\bm{Y'} + \bm{Y}\,&\bm{X'},\bm{c_2} = \bm{X}\,\bm{X'},\ \bm{c_3} = \bm{Y}\,\bm{Y'},
\end{align}
where the computation of $\bm{c_1}, \bm{c_2}, \bm{c_3}$ is subdue to the encrypted operator, controller and end\-/user communication:
\begin{align}
\enc_{\emph{pk1}}(\bm{c_1}) &= \enc_{\emph{pk1}}(\bm{Y})^{\bm{X'}}\circ\enc_{\emph{pk1}}(\bm{Y'})^{\bm{X}},\nonumber\\
\enc_{\emph{pk1}}(\bm{c_2} + \bm{c_3}) &= \enc_{\emph{pk1}}(\bm{X}\,\bm{X'}) \circ \enc_{\emph{pk1}}(\bm{Y}\,\bm{Y'}),
\end{align}
where $\circ$ denotes the Hadamard product\footnote{The Hadamard product is an entrywise product of two matrices $\bm{A}, \bm{B}$ having the same dimension: $\bm{A} \circ \bm{B} = (\bm{A})_{i,j}\,(\bm{B})_{i,j}$.}, and the terms $\enc_{\emph{pk1}}(\bm{Y})^{\bm{X'}}$, $\enc_{\emph{pk1}}(\bm{Y'})^{\bm{X}}$ represent exponentiations in an outer product fashion, resulting in the matrices $\enc_{\emph{pk1}}(\bm{Y}\,\bm{X'})$ and $\enc_{\emph{pk1}}(\bm{X}\,\bm{Y'})$, respectively.
Finally, the protected reference is $\bm{Y}^{\enc_{\emph{pk1}}}_{\emph{2Cov}} = \left(\enc_{\emph{pk1}}(\bm{Y}), \enc_{\emph{pk1}}(\bm{Y}\,\bm{Y'})\right)$.

\begin{figure}[htb]
    \centering
    \begin{tikzpicture}[font=\tiny, node distance=1em]
    \node [module step] 
        (step1a) {\stepno{1a.} \textbf{Extract probe:} \hfill\matharchitecturenode{$\bm{X}$}};
    \node [below=1.25em of step1a.south west,module step]
        (step1b) {\stepno{1b.} \textbf{Encrypt:} \\\hfill\matharchitecturenode{$\enc_{\emph{pk1}}\left(\bm{X}\,\bm{X'}\right)$}};
    \node [below=1.5em of step1b.south west,module step]
        (step3) {\stepno{3.} \textbf{Compute:}  \\\hfill\matharchitecturenode{$\enc_{\emph{pk1}}(\bm{c_1})$\\\hfill\matharchitecturenode{$={} \stepno{2a}^{\bm{X'}} \circ \stepno{2a}{\color{hda}'}^{\bm{X}}$}}};
    \node [below=1.25em of step3.south west,module step]
        (step4) {\stepno{4.} \textbf{Compute:} \\\hfill\matharchitecturenode{$\enc_{\emph{pk1}}(\bm{c_2\!+\!c_3})$\\\hfill\matharchitecturenode{$={}\stepno{1b}\circ\stepno{2b}$}}};
    \pic {module={Client}{step1a}{step4}};
    \coordinate (ClientEast) at ([yshift=1em]$(step1.north east)!0.5!(step4.south east)$);
    \coordinate (ClientSouth) at ($(step4.south west)$);
    \coordinate (DB1) at ([xshift=9em,yshift=-0.5em]ClientEast);
    \node [right=of DB1,module data]
        (data1) {\matharchitecturenode{$\enc_{\emph{pk1}}(\bm{Y})$}};
    \node [below=1.25em of data1.east,module data,anchor=east,yshift=0.0em]
        (data2) {\matharchitecturenode{$\enc_{\emph{pk1}}\left(\bm{Y}\,\bm{Y'}\right)$}};
    \pic {module={$\mathrm{DB}_{\emph{controller}}$}{data1}{data2}};
    \coordinate (Comm1) at ([xshift=-2em,yshift=0em]$(ClientEast)!0.5!(DB1)$);
    \node [at=(Comm1),channel] 
        (step2a) {\stepno{2a.} \textbf{Send:} \\\hfill\matharchitecturenode{$\enc_{\emph{pk1}}(\bm{Y})$}};
    \node [below=of step2a.south west,channel] 
        (step2b) {\stepno{2b.} \textbf{Send:} \\\hfil\matharchitecturenode{$\enc_{\emph{pk1}}\left(\bm{Y}\,\bm{Y'}\right)$}};
    \pic {channel left={Comm1}{2.75em}{-.85em}{7em}{7.5em}{.5em}{120}};
    \coordinate (AS1) at ([yshift=-9em]ClientSouth);
    \node [below=of AS1,module step] 
        (step7) {\stepno{7.} \textbf{Decrypt:} \\ \hfill\matharchitecturenode{$ \bm{c_1} = \dec_{\emph{sk1}}\left(\stepno{5a}\right)$}};
    \node [below=1.25em of step7.south west,module step] 
        (step8) {\stepno{8.} \textbf{Decrypt:} \\\hfill\matharchitecturenode{$\bm{c_2\!+\!c_3} = \dec_{\emph{sk1}}\left(\stepno{5b}\right)$}};
    \node [below=1.5em of step8.south west,module step] 
        (step9) {\stepno{9.} \textbf{Compute:} \\ \hfill\matharchitecturenode{$ \enc_{\emph{pk2}}\left(S_{\emph{2Cov}}\right)$}\\\hfill\matharchitecturenode{$ = {}\stepno{6a}^{\langle\rangle(\stepno{7})}\,\stepno{6b}^{\langle\rangle(\stepno{8})}$}};
    \node [below=of step9.south east,module data,anchor=east,yshift=0.375em]
        (dataASoperator) {\matharchitecturenode{${\emph{pk1}}, {\emph{sk1}}, {\emph{pk2}}$}};
    \pic {module={$\mathrm{AS}_{\emph{operator}}$}{step7}{dataASoperator}};
    \coordinate (ASEastNorth) at ([yshift=1.5em]$(step7.north east)$);
    \coordinate (ASEastSouth) at ([yshift=2.75em]$(step9.north east)!0.5!(dataASoperator.south east)$);
    \coordinate (Comm2) at ([xshift=2em,yshift=2.5em]ClientSouth|-AS1);
    \node [at=(Comm2),channel,rotate=90,xshift=0em] 
        (step5a) {\stepno{5a.} \textbf{Send:} \\\hfill\matharchitecturenode{$\enc_{\emph{pk1}}\!\left(\bm{c_1}\right)$}};
    \node [right=of step5a.south west,channel,rotate=90,xshift=0em] 
        (step5b) {\stepno{5b.} \textbf{Send:} \\\hfill\matharchitecturenode{$\enc_{\emph{pk1}}\!\left(\bm{c_2}\!+\!\bm{c_3}\right)$}};
    \pic {channel down={Comm2}{2.75em}{.85em}{7em}{7.5em}{.5em}{120}};
    \coordinate (DB2) at ([xshift=9em,yshift=-2em]ASEastNorth);
    \node [right=of DB2,module data]
        (data3) {\matharchitecturenode{$\enc_{\emph{pk2}}(\bm{\Lambda})$}};
    \node [below=1.25em of data3.east,module data,anchor=east,yshift=0.0em]
        (data4) {\matharchitecturenode{$\enc_{\emph{pk2}}(\bm{\Gamma})$}};
    \pic {module={$\mathrm{DB}_{\emph{vendor}}$}{data3}{data4}};
    \coordinate (Comm3) at ([xshift=-2em,yshift=-.5em]$(ASEastNorth)!0.5!(DB2)$);
    \node [at=(Comm3),channel] 
    (step6a) {\stepno{6a.} \textbf{Send:} \\\hfill\matharchitecturenode{$\enc_{\emph{pk2}}(\bm{\Lambda})$}};
    \node [below=of step6a.south west,channel] 
    (step6b) {\stepno{6b.} \textbf{Send:} \\\hfill\matharchitecturenode{$\enc_{\emph{pk2}}(\bm{\Gamma})$}};
    \pic {channel left={Comm3}{2.75em}{-.85em}{7em}{7.5em}{.5em}{120}};
    \coordinate (AS2) at ([xshift=10em,yshift=-2em]ASEastSouth);
    \node [below=of AS2,module step] 
        (step11) {\stepno{11.} \textbf{Decrypt score:} \\ \hfill\matharchitecturenode{$S_{\emph{2Cov}} = \dec_{\emph{sk2}}\left(\stepno{10}\right)$}};
    \node [below=of step11.south west,module step] 
        (step12) {\stepno{12.} \textbf{Decision:} \\ \hfill\matharchitecturenode{$(\eta \leq \stepno{11}) \rightarrow$ \textbf{yes\,/\,no}}};
    \node [below=of step12.south east,module data,anchor=east,yshift=0.375em]
        (dataASvendor) {\matharchitecturenode{${\emph{pk2}}, {\emph{sk2}}$}};
    \pic {module={$\mathrm{AS}_{\emph{vendor}}$}{step11}{dataASvendor}};
    \coordinate (Comm4) at ([yshift=-2.0em]Comm3|-AS2);
    \node [at=(Comm4),channel,rotate=0,xshift=0em] 
        (step10) {\stepno{10.} \textbf{Send:} \\\hfill\matharchitecturenode{$\enc_{\emph{pk2}}\left(S_{\emph{2Cov}}\right)$}};
    \pic {channel right={Comm4}{2.25em}{0em}{7em}{4.25em}{.5em}{120}};
\end{tikzpicture}
    \caption{Architecture of protected templates and hyper-parameters, with client, servers (blue) and communication channels (orange).}
    \label{fig:HE:2Cov_schema_subject_vendor}
\end{figure}
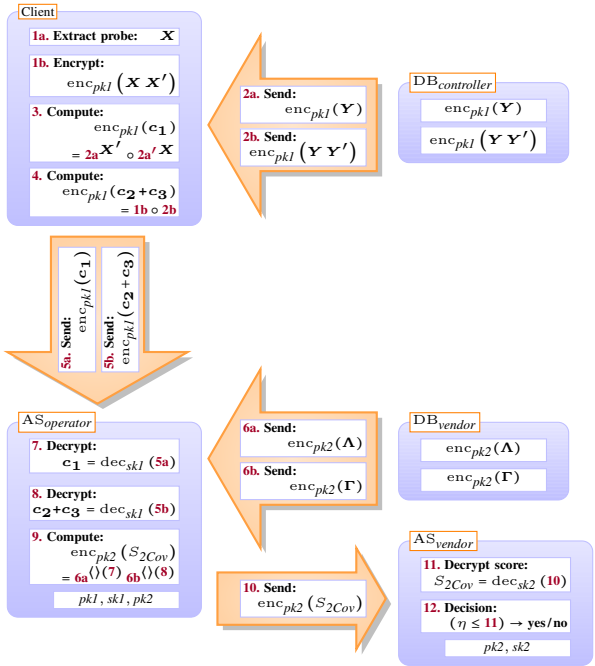

Fig.~\ref{fig:HE:2Cov_schema_subject_vendor} presents the proposed architecture.
The previously proposed architecture is extended by additional communication channels between operators and vendors.
Applications employ two key pairs, such that template protection can be achieved dependent on both: (a) different biometric services of an operator, and (b) multiple provisions of a biometric system to service operators by a vendor.
Consequently, additional servers are necessary on the vendor site in terms of a database $\mathrm{DB}_{\emph{vendor}}$ and an authentication server $\mathrm{AS}_{\emph{vendor}}$, respectively.

\section{Experimental Analysis and Discussion}
\label{sec:experiments}

\begin{figure}[htb]
    \centering
    \setvariable{\detsize}{175pt}
    \input{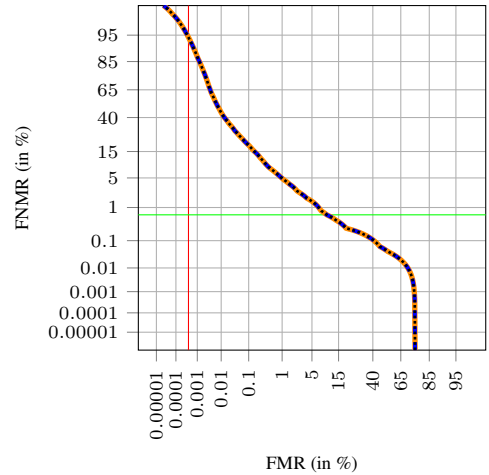}
    \caption{DET comparison of the baseline 2Cov system (orange), and the proposed HE 2Cov schemes, focusing on subject data protection (blue, dashed), and the protection of subject and vendor data (black, dotted) with rule of 30 bounds (red, green).}
    \label{fig:HE:2covPerformance}
\end{figure}

An experimental validation is conducted on the 2013\,--\,2014 NIST i\-/vector machine learning challenge \cite{NIST-2013-2014-Speaker-Recognition-i-vector-Challenge-NIST-2013,Banse-NIST-SRE-2014-Interspeech-2014} phase III database (\ie with labeled development data), where 600 dimensional i\-/vectors are supplied, comprising a development set of $36\,572$ i\-/vectors, $1\,306$ references with each five enrolment \mbox{i\-/vectors}, and $9\,634$ probes, conducting $12\,582\,004$ comparisons on averaged reference i\-/vectors as template models.
The prototype system comprises a dimension reduction to $F=250$ by linear discriminant analysis, within class covariance normalization, length normalization, and 2Cov comparison.
For the Paillier cryptosystem, $n=2\,048\bits$ keys are utilized, in accordance with the NIST recommendation \cite{Barker-PairWiseKeyRecommendation-NIST-2013}.
In contrast, plaintext operations consider double floating\-/point precision, \ie $p=64\bits$ per plain real feature value.
Implementations are based on the freely available \emph{sidekit} \cite{Larcher-SIDEKIT-extensible-speaker-identification-Python-ICASSP-2016} and \emph{Python Paillier} \cite{Thorne-PythonPaillier-2017}.
Fig.~\ref{fig:HE:2covPerformance} illustrates the DET performance of conventional and HE 2Cov comparators on the evaluation set in terms of false non\-/match rate (FNMR) and false match rate (FMR): the baseline performance is preserved across all operating points.
The DET is depicted in terms of the convex hull of the receiver operating characteristic (ROCCH).
For the exemplary 2Cov system, a $2.5\%$ equal error rate (ROCCH\-/EER), a $0.050$ minDCF (parameterized according to \cite{NIST-2013-2014-Speaker-Recognition-i-vector-Challenge-NIST-2013}), and a $0.099$ \Cllrmin are preserved in the protected domain.
As the $k$ normalization term is neglected in this set-up, the baseline system yielded a $9.560$ \Cllr.
Calibration loss can be reduced by a post score re-bias, or by employing conventional score calibration methods, \cf \cite{Brummer-deVilliers-BOSARIS-Binary-Scores-AGNITIO-Research-2011}.
By utilizing linear score calibration trained on the oracle development set, \Cllr is reduced to $0.284$.


As the verification performance is preserved, the proposed schemes are further examined regarding requirements of the biometric template protection standard \cite{ISO-IEC-24745-TemplateProtection-101129} in terms of \cite{gomez2016implementation}: \emph{$i)$ only the client can have access to the plain probe template, $ii)$ the plain reference template should not be seen by the client, and only its encryption should be stored or handled during verification, and $iii)$ the score should also be protected in order to prevent hill\-/climbing and inverse\-/biometrics attacks}.
Firstly, both employed homomorphic Paillier cryptosystem provide semantic security: only secret keys are able to derive the plain probe after encryption, where the client solely communicates the encrypted score $\left(\enc_{pk}\left(S_{\emph{2Cov}}\left(\bm{X}, \bm{Y}\right)\right)\right)$ or auxiliary matrices $\left(\enc_{pk1}\left(\bm{c_1}\right), \enc_{pk1}\left(\bm{c_2} + \bm{c_3}\right)\right)$.
Secondly, biometric references are communicated from the controller database server to the client in the encrypted domain, assuming the authentication server being able to protect the secret key $sk1$, no other entities will be able to relate the protected biometric information.
Similarly, the vendor data is protected in the sense, that the vendor authentication server is assumed to be able to protect $sk2$.
Finally, scores are computed in the protected domain, and can solely be decrypted utilizing secret key $sk2$.
Thus, the irreversibility criterion is met.
Renewability is granted as depicted in \cite{gomez2016implementation}: if templates are lost, new key pairs can be generated for the purpose of re\-/encrypting the database, such that (a) re\-/acquisitions of enrollment samples are avoided when revoking corrupted templates, and (b) comparisons of corrupt to renewed templates result in non\-/matches, granting security and data privacy.
Thus, templates can easily be revoked, thereby providing data privacy.
Unlinkability is granted due to the probabilistic nature of the Paillier cryptosystem, where random numbers are utilized for different encryptions, \ie encrypting the same data $\bm{Y}$ twice, two different random numbers $s_1, s_2$ are drawn, such that: $\enc_e(\bm{Y}, s_1) \neq \enc_e(\bm{Y}, s_2)$, \cf \cite{gomez2016implementation,paillier1999public}.

\begin{table}[htb]
    \caption{Complexity analysis for the proposed 2Cov HE schemes (verification) with the data sizes of the exemplary employed system ($p=64\bits, \nu=0.5\KiB, F=250$).}
    \label{tab:complexity}
    \scriptsize
    \centering
    \smallskip
    \begin{tabular}{lcc} 
        \toprule
        Comparator & 2Cov & 2Cov \\ 
        Protection & subject & subject \& vendor\\
        \midrule
        N$^o$ encryptions & $1$ & $F^2$ \\
        N$^o$ decryptions & $1$ & $2\,F^2 + 1$ \\
        \midrule
        N$^o$ additions & $4\,F\,(F-1)$ & $0$ \\
        N$^o$ products & $4\,F^2 + 2\,F + 1$ & $5\,F^2 - 1$\\
        N$^o$ exponentiations & $2\,F$ & $4\,F^2$\\
        \midrule
        Plain template size & $p\,F$ & $p\,F$ \\
        & $\approx 2.0\KiB$ & $\approx 2.0\KiB$ \\
        Protected template size & $\nu\,(F+1)$ & $\nu\,(F^2 + F)$ \\
        & $=125.5\KiB$ & $\approx 30.6\MiB$\\
        \midrule
        Plain model size & $2\,p\,F^2$ & $2\,p\,F^2$ \\
        & $\approx 1.0\MiB$ & $\approx 1.0\MiB$\\
        Protected model size & $0$ & $2\,\nu\,F^2$ \\
        & $=0\KiB$ & $\approx61.0\MiB$ \\
        \midrule
        Channels: amount of & $\nu\,(F+2)$ & $\nu\,(5\,F^2 + F + 1)$ \\
        protected data exchanged & $= 126.0\KiB$ & $\approx 152.7\MiB$\\
        \bottomrule
    \end{tabular}
\end{table}

In terms of complexity, each approach can be analyzed regarding the amount of required resources, \ie the number of operations performed in the encrypted domain as well as the size of encrypted data sent over a channel.
For a single verification attempt, the chipertext channel bandwidth is $\nu = 2\,n$ due to the Paillier ciphertext length in modulo $n^{2}$ domain \cite{gomez2016implementation}, \ie $\nu = 4\,096\bits\,\frac{1\KiB}{8\,192\bits} = 0.5\KiB$ for the examined system.
Tab.~\ref{tab:complexity} summarizes the proposed HE schemes' complexity.
Regarding to an i\-/vector dimension $F=250$, the cosine HE approach requires $\nu\,F = 125\KiB$ for storing a reference i\-/vector.
For transmitting the protected score to the authentication server, $0.5\KiB$ are necessary, \ie a protected scalar.
The subject protective 2Cov HE scheme stores a reference tuple with $\nu\,(F+1) = 125.5\KiB$, communicating a protected scalar as well to the authentication server.
However, the subject and vendor protective 2Cov HE scheme stores protected auxiliary matrices, requiring $\nu\,(F^2 + F) \approx 30.6\MiB$.
Therefore, the channel between client and authentication server considers two protected matrices, requiring $2\,\nu\,F^2 \approx 61.0\MiB$, alike for the vendor database to operator authentication server channel.
Regarding the protected data exchanged over the communication channels, the first proposed scheme comprises $\nu\,(F+2) = 126\KiB$ as the protected template and score are transmitted.
The second proposed scheme demands higher requirements: as the model hyper-parameters are protected, the client to authentication server channel transmits auxiliary matrices comprising $2\,\nu\,(F^2)\approx61.0\MiB$, whereas the same data amount is loaded for the protected model from the vendor database server.
Finally, a protected score is transmitted to the vendor authentication server, making application decisions.
Afterwards, conventional security protocols can be employed.
\section{Conclusion}
\label{sec:conclusion}

Homomorphic template protection is made available to generative comparators, \ie comparators employing statistical models, where the related biometrics work solely considers non\-/generative comparators, such as XOR, DTW, Euclidean distance, and cosine similarity.
Extending the HE scheme for cosine similarity comparison, template protection is made available to the 2Cov comparator in two architectures.
The first proposed HE architecture solely puts emphasis on the protection of templates, which can be sustained under a fair complexity tradeoff.
Contrastively, the second proposed HE 2Cov scheme provides subject and vendor data protection.
However, the required complexity increases by a quadratic term.
By pre\-/loading both protected model parameters the channel bottleneck is reduced to $\nu\,(3\,F^2 + F + 1) \approx 91.7\MiB$ for a single verification attempt, which however limits the application scope to well\-/equipped infrastructures \eg, call center and forensic scenarios.
Depending on the application scenario, protected templates may also be pre\-/loaded, further reducing the overall transmitted data amount to $\nu\,(2\,F^2 + 1) \approx 61.0 \MiB$.
For mobile device voice biometrics, one may prefer to employ the first proposed architecture.
Both approaches ensure biometric template protection requirements as of the ISO/IEC~24745 standard.
For the sake of reproducibility, we provide a reference implementation.

As the proposed schemes target 2Cov as prototype generative comparators, \ie the full\-/subspace Gaussian PLDA special case, extensions to other members of the PLDA family and related comparators can be easily developed.
Accounting for \mbox{i\-/vectors} not only as single point estimate features but also as latent variables, uncertainties associated to the single point estimation can be incorporated as well, \eg targeting full\-/posterior PLDA.
Also, HE schemes seem promising for end\-/to\-/end neural network system architectures, as the inner Frobenius product is computable in the protected domain.
Extensions of the proposed architectures and implementations of alternative HE schemes is left to future work.

\section{Acknowledgements}
This work was supported by the German Federal Ministry of Education and Research (BMBF) as well as by the Hessen State Ministry for Higher Education, Research and the Arts within the Center for Research in Security and Privacy (\url{www.crisp-da.de}), and the BioMobile II project (no. 518/16-30).

\vfill
\newpage

\bibliographystyle{IEEEbib}
\bibliography{biometrics,thesis,jascha}

\begin{thebibliography}{10}

\bibitem{EU-Directive-DataPrivacy-160427}
{European Council},
\newblock ``Directive 2016/680 of the {E}uropean {P}arliament and of the
  {C}ouncil on the protection of individuals with regard to the processing of
  personal data by competent authorities for the purposes of the prevention,
  investigation, detection or prosecution of criminal offences or the execution
  of criminal penalties, and on the free movement of such data, and repealing
  {C}ouncil {F}ramework {D}ecision {2008/977/JHA},'' April 2016.

\bibitem{EU-Directive-PSD2-151125}
{European Parliament} and {European Council},
\newblock ``Directive 2015/2366 of the {E}uropean {P}arliament and of the
  {C}ouncil of 25 {N}ovember 2015 on payment services in the internal market,''
  November 2015.

\bibitem{ISO-IEC-24745-TemplateProtection-101129}
{ISO/IEC JTC1 SC27 Security Techniques},
\newblock {\em {ISO/IEC} 24745:2011. Information Technology - Security
  Techniques - Biometric Information Protection},
\newblock International Organization for Standardization, 2011.

\bibitem{Vaquero-VoiceBTP-Interspeech-2015}
C.~Vaquero and P.~Rodr{\'i}guez,
\newblock ``On the need of template protection for voice authentication,''
\newblock in {\em Proc. Annual Conf. of the Intl. Speech Communication
  Association (INTERSPEECH)}, 2015, pp. 219--223.

\bibitem{Cappelli07PAMIReconstruction}
R.~Cappelli, D.~Maio, A.~Lumini, and D.~Maltoni,
\newblock ``Fingerprint image reconstruction from standard templates,''
\newblock {\em IEEE Trans. on Pattern Analysis and Machine Intelligence}, vol.
  29, no. 9, pp. 1489--1503, 2007.

\bibitem{Galbally2013}
J.~Galbally, A.~Ross, M.~Gomez-Barrero, J.~Fierrez, and J.~Ortega-Garcia,
\newblock ``Iris image reconstruction from binary templates: An efficient
  probabilistic approach based on genetic algorithms,''
\newblock {\em Computer Vision and Image Understanding}, vol. 117, no. 10, pp.
  1512--1525, 2013.

\bibitem{Glembek-MigratingIvectors-Interspeech-2015}
O.~Glembek, P.~Matejka, O.~Plchot, J.~Pesan, L.~Burget, and P.~Schwarz,
\newblock ``Migrating i-vectors between speaker recognition systems using
  regression neural networks,''
\newblock in {\em Proc. Annual Conf. of the Intl. Speech Communication
  Association (INTERSPEECH)}, 2015, pp. 2327--2331.

\bibitem{Moyakine-SIIPPanelEthics-EABrpc-2017}
E.~Moyakine, C.~Colonnello, J.~Butler, and C.~Jasserand,
\newblock ``Discussion panel: {SIIP} and {INGRESS} research projects:
  Developing effective and sustainable biometric systems with a global reach,''
  2017,
\newblock EAB Research Projects Conference, [Online]
  \url{https://www.eab.org/upload/documents/1279/08-eabrpc2017_SIIP_INGRESS.zip?ts=1517990107115},
  last accessed: 2018-02-07.

\bibitem{Rathgeb-BTP-Survey-EURASIP-2011}
C.~Rathgeb and A.~Uhl,
\newblock ``A survey on biometric cryptosystems and cancelable biometrics,''
\newblock {\em EURASIP Journal on Information Security}, vol. 3, 2011.

\bibitem{patel15CancelableBioSurvey}
V.~M. Patel, N.~Ratha, and R.~Chellappa,
\newblock ``Cancelable biometrics: A review,''
\newblock {\em IEEE Signal Proc. Magazine}, vol. 32, no. 5, pp. 54--65, 2015.

\bibitem{campisi13secPrivacyBio}
P.~Campisi, Ed.,
\newblock {\em Security and Privacy in Biometrics},
\newblock Springer, 2013.

\bibitem{fontaine13surveyHE}
C.~Aguilar-Melchor, S.~Fau, C.~Fontaine, et~al.,
\newblock ``Recent advances in homomorphic encryption: A possible future for
  signal processing in the encrypted domain,''
\newblock {\em IEEE Signal Processing Magazine}, vol. 30, no. 2, pp. 108--117,
  2013.

\bibitem{gomez2016implementation}
M.~Gomez-Barrero, J.~Fierrez, J.~Galbally, E.~Maiorana, and P.~Campisi,
\newblock ``Implementation of fixed length template protection based on
  homomorphic encryption with application to signature biometrics,''
\newblock in {\em Proc. Conf. on Computer Vision and Pattern Recognition
  Workshops (CVPR)}, 2016, pp. 191--198.

\bibitem{GomezBarrero-BiometricTemplateProtection-InformationScience-2016}
M.~Gomez-Barrero, C.~Rathgeb, J.~Galbally, C.~Busch, and J.~Fierrez,
\newblock ``Unlinkable and irreversible biometric template protection based on
  bloom filters,''
\newblock {\em Information Sciences}, vol. 370--371, pp. 18--32, 2016.

\bibitem{GomezBarrero-MBTPwithHE-PR-2017}
M.~Gomez-Barrero, E.~Maiorana, J.~Galbally, P.~Campisi, and J.~Fierrez,
\newblock ``Multi-biometric template protection based on {Homomorphic
  Encryption},''
\newblock {\em Pattern Recognition}, vol. 67, pp. 149--163, 07 2017.

\bibitem{GomezBarrero-VariableLengthBTPHE-Access-2017}
M.~Gomez-Barrero, J.~Galbally, A.~Morales, and J.~Fierrez,
\newblock ``Privacy-preserving comparison of variable-length data with
  application to biometric template protection,''
\newblock {\em IEEE Access}, vol. 5, no. 1, pp. 8606--8619, 12 2017.

\bibitem{Cumani-PairwiseDiscriminativeVerification-TASLP-2013}
S.~Cumani, N.~Br{\"u}mmer, L.~Burget, P.~Laface, O.~Plchot, and V.~Vasilakakis,
\newblock ``Pairwise discriminative speaker verification in the i-vector
  space,''
\newblock {\em IEEE Trans. on Audio, Speech, and Language Processing (TASLP)},
  vol. 21, no. 6, pp. 1217--1227, 2013.

\bibitem{Cumani-GenerativePairwiseModels-Odyssey-2014}
S.~Cumani and P.~Laface,
\newblock ``Generative pairwise models for speaker recognition,''
\newblock in {\em Proc. Odyssey 2014: The Speaker and Language Recognition
  Workshop}, 2014, pp. 273--279.

\bibitem{paillier1999public}
P.~Paillier,
\newblock ``Public-key cryptosystems based on composite degree residuosity
  classes,''
\newblock in {\em Proc. Advances in Cryptology --- EUROCRYPT}, 1999, pp.
  223--238.

\bibitem{Zhu-PrivacyPreservingTimeSeriesData-EDBT-2014}
H.~Zhu, X.~Meng, and G.~Kollios,
\newblock ``Privacy preserving similarity evaluation of time series data,''
\newblock in {\em Proc. Intl. Conf. on Extending Database Technology (EDBT)},
  2014, pp. 499--510.

\bibitem{penn2014customisation}
G.~M. Penn, G.~P{\"o}tzelsberger, M.~Rohde, and A.~Uhl,
\newblock ``Customisation of paillier homomorphic encryption for efficient
  binary biometric feature vector matching,''
\newblock in {\em Proc. IEEE Intl. Conf. Biometrics Special Interest Group
  (BIOSIG)}, 2014, pp. 1--6.

\bibitem{barniprivacy}
M.~Barni, T.~Bianchi, D.~Catalano, M.~{Di Raimondo}, R.~D. Labati, et~al.,
\newblock ``A privacy-compliant fingerprint recognition system based on
  homomorphic encryption and fingercode templates,''
\newblock in {\em Proc. Intl. Conf. on Biometrics: theory applications and
  systems (BTAS)}. IEEE, 2010, pp. 1--7.

\bibitem{Dehak-FrontEndFactorAnalysis-ivector-TASLP-2011}
N.~Dehak, P.~J. Kenny, R.~Dehak, P.~Dumouchel, and P.~Ouellet,
\newblock ``Front-end factor analysis for speaker verification,''
\newblock {\em IEEE Trans. on Audio, Speech, and Language Processing (TASLP)},
  vol. 19, no. 4, pp. 788--798, 2011.

\bibitem{Kenny-JFA-Theory-CRIM-2005}
P.~Kenny,
\newblock ``Joint factor analysis of speaker and session variability: Theory
  and algorithms,''
\newblock Tech. {R}ep. CRIM-06/08-13, CRIM, Montreal, 2005.

\bibitem{AngueraBonastre-SpeakerBinaryKey-Interspeech-2010}
X.~Anguera and J.~F. Bonastre,
\newblock ``A novel speaker binary key derived from anchor models,''
\newblock in {\em Proc. Annual Conf. of the Intl. Speech Communication
  Association (INTERSPEECH)}, 2010, pp. 2118--2121.

\bibitem{Bonastre-DiscriminantBinaryData-ICASSP-2011}
J.~F. Bonastre, P.~M. Bousquet, D.~Matrouf, and X.~Anguera,
\newblock ``Discriminant binary data representation for speaker recognition,''
\newblock in {\em Proc. IEEE Intl. Conf. on Acoustics, Speech, and Signal
  Processing (ICASSP)}, 2011, pp. 5284--5287.

\bibitem{Hernandez-SpeakerBinaryRepresentation-ProgressPatternRecognition-2012}
G.~Hern{\'a}ndez-Sierra, J.~F. Bonastre, and J.~{Calvo de Lara},
\newblock ``Speaker recognition using a binary representation and specificities
  models,''
\newblock in {\em Proc. Iberoamerican Congress on Pattern Recognition (CIARP)},
  2012, pp. 732--739.

\bibitem{Paulini-MultiBitAllocation-Odyssey-2016}
M.~Paulini, C.~Rathgeb, A.~Nautsch, H.~Reichau, H.~Reininger, and C.~Busch,
\newblock ``Multi-bit allocation: Preparing voice biometrics for template
  protection,''
\newblock in {\em Proc. Odyssey 2016: The Speaker and Language Recognition
  Workshop}, 2016, pp. 291--296.

\bibitem{Barni-BTP-SPM-2015}
M.~Barni, G.~Droandi, and R.~Lazzeretti,
\newblock ``Privacy protection in biometric-based recognition systems: A
  marriage between cryptography and signal processing,''
\newblock {\em IEEE Signal Processing Magazine}, vol. 32, no. 5, pp. 66--76,
  2015.

\bibitem{Bringer-BTP-Survey-SPM-2013}
J.~Bringer, H.~Chabanne, and A.~Patey,
\newblock ``Privacy-preserving biometric identification using secure multiparty
  computation: An overview and recent trends,''
\newblock {\em IEEE Signal Processing Magazine}, vol. 30, no. 2, pp. 42--52,
  2013.

\bibitem{Jain-Fingercodes-IEEE-1997}
A.~K. Jain, L.~Hong, S.~Pankanti, and R.~Bolle,
\newblock ``An identity-authentication system using fingerprints,''
\newblock {\em Proc. of the IEEE}, vol. 85, no. 9, pp. 1365--1388, 1997.

\bibitem{Bianchi-BTP-Fingercode-BIOMS-2010}
T.~Bianchi, S.~Turchi, A.~Piva, R.~D. Labati, V.~Piuri, and F.~Scotti,
\newblock ``Implementing fingercode-based identity matching in the encrypted
  domain,''
\newblock in {\em IEEE Workshop on Biometric Measurements and Systems for
  Security and Medical Applications {(BIOMS)}}, 2010, pp. 15--21.

\bibitem{Ye-BTP-HE-EURASIP-2009}
S.~Ye, Y.~Luo, J.~Zhao, and S.~Cheung,
\newblock ``Anonymous biometric access control,''
\newblock {\em EURASIP Journal on Information Security}, vol. 2009, no. 1, pp.
  1--17, 2009.

\bibitem{Luo-BTP-HE-ICME-2009}
Y.~Luo, S.~C. Sen-ching, and S.~Ye,
\newblock ``Anonymous biometric access control based on homomorphic
  encryption,''
\newblock in {\em Proc. Intl. Conf. on Multimedia and Expo (ICME)}, 2009, pp.
  1046--1049.

\bibitem{Blanton-BTP-IrisFingerprint-ESORICS-2011}
M.~Blanton and P.~Gasti,
\newblock {\em Secure and Efficient Protocols for Iris and Fingerprint
  Identification}, pp. 190--209,
\newblock Springer Berlin Heidelberg, 2011.

\bibitem{quantum}
D.~J. Bernstein, J.~Buchmann, and E.~Dahmen,
\newblock {\em Post-Quantum Cryptography},
\newblock Springer Science \& Business Media, 2009.

\bibitem{Yasuda-BTP-HE-ARES-2013}
M.~Yasuda, T.~Shimoyama, J.~Kogure, K.~Yokoyama, and T.~Koshiba,
\newblock ``Packed homomorphic encryption based on ideal lattices and its
  application to biometrics,''
\newblock in {\em Proc. Intl. Conf. on Availability, Reliability, and
  Security}, 2013, pp. 55--74.

\bibitem{Yasuda-BTP-HE-SCN-2015}
M.~Yasuda, T.~Shimoyama, J.~Kogure, K.~Yokoyama, and T.~Koshiba,
\newblock ``New packing method in somewhat homomorphic encryption and its
  applications,''
\newblock {\em Security and Communication Networks}, vol. 8, no. 13, pp.
  2194--2213, 2015.

\bibitem{Patsakis-BTP-HE-DPM-2015}
C.~Patsakis, van J.~Rest, M.~Chora{\'s}, and M.~Bouroche,
\newblock ``Privacy-preserving biometric authentication and matching via
  lattice-based encryption,''
\newblock in {\em Proc. Intl. Workshop on Data Privacy Management}, 2015, pp.
  169--182.

\bibitem{Rivest-PrivacyHomomorphisms-SecureComputation-1978}
R.~L. Rivest, L.~Adleman, and M.~L. Dertouzos,
\newblock ``On data banks and privacy homomorphisms,''
\newblock {\em Foundations of secure computation}, vol. 4, no. 11, pp.
  168--180, 1978.

\bibitem{FontaineGaland-Survey-HomomorphicEncryption-JIS-2007}
C.~Fontaine and F.~Galand,
\newblock ``A survey of homomorphic encryption for nonspecialists,''
\newblock {\em EURASIP Journal on Information Security}, vol. 2007, 2007.

\bibitem{Hoffstein-IntroductionCryptography-2008}
J.~Hoffstein, J.~Pipher, and J.~H. Silverman,
\newblock {\em An Introduction to Mathematical Cryptography},
\newblock Springer, 2008.

\bibitem{Hungerford-Algebra-Springer-1974}
T.~W. Hungerford,
\newblock {\em Algebra},
\newblock Springer Graduate Texts in Mathematics, 1974.

\bibitem{bellare1998relations}
M.~Bellare, A.~Desai, D.~Pointcheval, and P.~Rogaway,
\newblock ``Relations among notions of security for public-key encryption
  schemes,''
\newblock in {\em Proc. Advances in Cryptology (CRYPTO)}, 1998, pp. 26--45.

\bibitem{paillier1999efficient}
P.~Paillier and D.~Pointcheval,
\newblock ``Efficient public-key cryptosystems provably secure against active
  adversaries,''
\newblock in {\em Proc. Advances in Cryptography --- ASIACRYPT}, 1999, pp.
  165--179.

\bibitem{Reynolds-GMMs-ConversationalSpeech-2000}
D.~A. Reynolds, T.~F. Quatieri, and R.~B. Dunn,
\newblock ``Speaker verification using adapted gaussian mixture models,''
\newblock {\em Conversational Speech, Digital Signal Processing}, vol. 10, pp.
  19--41, 2000.

\bibitem{GarciaRomeroEpsyWison-ivector-lengthnormalization-Interspeech-2011}
D.~Garcia-Romero and C.Y. Epsy-Wilson,
\newblock ``Analysis of i-vector length normalization in speaker recognition
  systems,''
\newblock in {\em Proc. Annual Conf. of the Intl. Speech Communication
  Association (INTERSPEECH)}, 2011, pp. 249--252.

\bibitem{Bousquet-BenefitsDifferentSteps-ivectors-CIARP-2013}
P.-M. Bousquet, J.-F. Bonastre, and D.~Matrouf,
\newblock {\em Identify the Benefits of the Different Steps in an i-Vector
  Based Speaker}, chapter CIARP, Part II, pp. 278--285,
\newblock Springer-Verlag Berlin Heidelberg, 2013.

\bibitem{ieee2008754}
{IEEE Standards Association},
\newblock {\em 754-2008 {IEEE} standard for Floating-Point Arithmetic}, 2008.

\bibitem{Thorne-PythonPaillier-2017}
B.~Thorne,
\newblock ``{P}ython {P}aillier,'' 2017,
\newblock [Online] \url{https://github.com/n1analytics/python-paillier/}, last
  accessed: 2018-01-11.

\bibitem{NIST-2013-2014-Speaker-Recognition-i-vector-Challenge-NIST-2013}
National~Institute of~Standards and Technology (NIST),
\newblock ``The 2013-2014 speaker recognition i-vector machine learning
  challenge,''
\newblock Tech. {R}ep., National Institute of Standards and Technology, 2014.

\bibitem{Banse-NIST-SRE-2014-Interspeech-2014}
D.~Bans{\'e}, G.~R. Doddington, D.~Garcia-Romero, J.~J. Godfrey, C.~S.
  Greenberg, et~al.,
\newblock ``Summary and initial results of the 2013-2014 speaker recognition
  i-vector machine learning challenge,''
\newblock in {\em Proc. Annual Conf. of the Intl. Speech Communication
  Association (INTERSPEECH)}, 2014, pp. 368--372.

\bibitem{Barker-PairWiseKeyRecommendation-NIST-2013}
E.~Barker, L.~Chen, A.~Roginsky, and M.~Smid,
\newblock ``Recommendation for pair-wise key establishment schemes using
  discrete logarithm cryptography,''
\newblock Tech. {R}ep. SP 800-56A Rev. 2, NIST, May 2013.

\bibitem{Larcher-SIDEKIT-extensible-speaker-identification-Python-ICASSP-2016}
A.~Larcher, K.A. Lee, and S.~Meignier,
\newblock ``An extensible speaker identification {SIDEKIT} in {P}ython,''
\newblock in {\em Proc. IEEE Intl. Conf. on Acoustics, Speech and Signal
  Processing ({ICASSP})}, 2016, pp. 5095--5099,
\newblock \url{http://lium.univ-lemans.fr/sidekit}, last accessed: 2017-05-15.

\bibitem{Brummer-deVilliers-BOSARIS-Binary-Scores-AGNITIO-Research-2011}
N.~Br{\"u}mmer and E.~{de Villiers},
\newblock ``The {BOSARIS} toolkit user guide: Theory, algorithms and code for
  binary classifier score processing,''
\newblock Tech. {R}ep., AGNITIO Research, South Africa, December 2011.

\end{thebibliography}

\end{document}